\documentclass[conference]{IEEEtran}
\IEEEoverridecommandlockouts
\usepackage{cite}
\usepackage{amsmath,amssymb,amsfonts}
\usepackage{algorithmic}
\usepackage{graphicx}
\usepackage{textcomp}
\usepackage{xcolor}
\usepackage{hyperref}
\usepackage{booktabs}
\usepackage{multirow}
\usepackage{graphicx}
\def\BibTeX{{\rm B\kern-.05em{\sc i\kern-.025em b}\kern-.08em
    T\kern-.1667em\lower.7ex\hbox{E}\kern-.125emX}}

\begin{document}

\title{Classification and understanding of cloud structures via satellite images with EfficientUNet\\
}

\author{\IEEEauthorblockN{
Tashin Ahmed}
tashinahmed@aol.com
\and
\IEEEauthorblockN{
Noor Hossain Nuri Sabab}
nsabab@aol.com
}

\maketitle

\begin{abstract}

Climate change has been a common interest and the forefront of crucial political discussion and decision-making for many years. Shallow clouds play a significant role in understanding the Earth's climate, but they are challenging to interpret and represent in a climate model. By classifying these cloud structures, there is a better possibility of understanding the physical structures of the clouds, which would improve the climate model generation, resulting in a better prediction of climate change or forecasting weather update. Clouds organise in many forms, which makes it challenging to build traditional rule-based algorithms to separate cloud features. In this paper, classification of cloud organization patterns was performed using a new scaled-up version of Convolutional Neural Network (CNN) named as EfficientNet as the encoder and UNet as decoder where they worked as feature extractor and reconstructor of fine grained feature map and was used as a classifier, which will help experts to understand how clouds will shape the future climate. By using a segmentation model in a classification task, it was shown that with a good encoder alongside UNet, it is possible to obtain good performance from this dataset. Dice coefficient has been used for the final evaluation metric, which gave the score of 66.26\% and 66.02\% for public and private (test set) leaderboard on Kaggle competition respectively.

\end{abstract}

\begin{IEEEkeywords}
EfficientNet, UNet, clouds, Dice coefficient
\end{IEEEkeywords}

\section{Introduction}

Clouds perform an essential role to control the radiation of the sun as well as controlling the radiation that goes back to the atmosphere. The more energy that is trapped inside the planet, the warmer the atmosphere becomes, giving rise to sea level via meltdown of polar ice caps and contributing to global warming. The less energy that is trapped, the colder the temperature becomes. Understanding the structure of the clouds gives a better insight into the planet's weather. Hence it is crucial to climatologists \cite{turner2007thin}. \textbf{Albedo} is a measure of how much energy is reflected without being absorbed \cite{twomey1974pollution}. White surfaces reflect the most energy; hence it has a high albedo, while dark surfaces absorb most energy, indicating a low albedo. Earth's albedo is 0.3, which indicate the warming of the climate \cite{palle2004changes}. Interpreting cloud structures provide useful insight into the abuse of Earth's climate and risks associated with it. Satellite images of clouds give a broader picture of the atmosphere, and interpreting the images provide information on the current situation of the planet.

\begin{figure}[ht!]
    \centering
    \includegraphics[width=\linewidth]{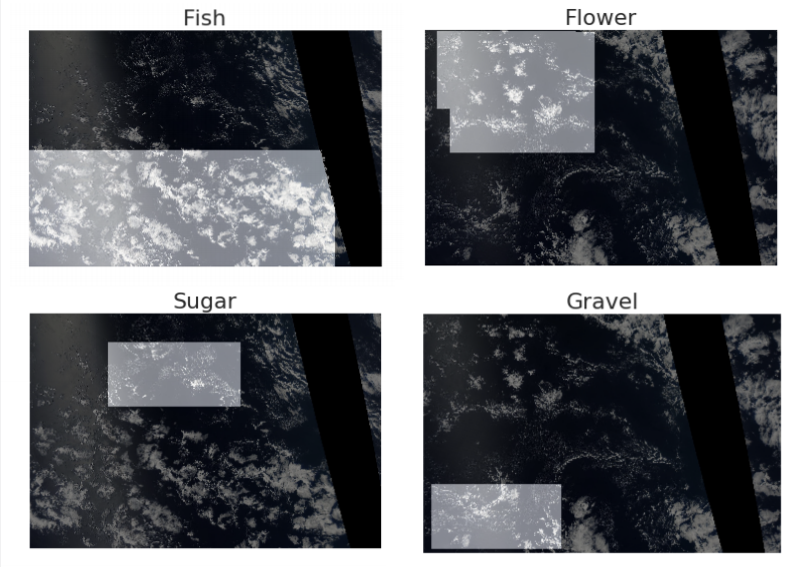}
    \caption{Masked images of four different class samples}
    \label{Masked}
\end{figure}

It is assumed that as the overall temperature of Earth increases, it will evaporate more water from the oceans, resulting in more clouds with different structures and variations \cite{leconte2013increased}. Climate change could depend on the general effect of clouds shapes and other features like more or less abundant, thicker or thinner and altitude positions. There are several types of cloud structures. Out of all top level clouds \textbf{Cirrus} clouds are the most abundant. Cirrus means a "curl of hair" \cite{dowling1990summary}.

These feathery clouds are made of icicles and have lengthy narrow strips that are also known as mare’s tails. Dispersed cirrus clouds indicates good weather. A continuous growth of a cirrus cloud which looks like a cover of the web indicates increasing humidity in air and arrival of a potential storm. \textbf{Cirrostratus} clouds lay themselves out across the sky with their scattered thin lines which gives the air a subtle white look. These are often the signs of upcoming rain. Their translucency offers the sun and moon to be seen easily through the cloud. Visibility of a cirrostratus cloud normally can be confirmed twelve to twenty four hours before the arrival of a rain or snow \cite{gadsden1989noctilucent}. \textbf{Cirrocomulus} is another type of high cloud which tends to be a cluster of white lines that oftentimes can be seen as orderly aligned. In tropical climates these could be the mark of an incoming hurricane \cite{mclean1957cloud}. Transverse cirrus bands could provide turbulence in planes and the heat maps annually show the spatial patterns where the bands are quite consistently forming \cite{miller2018detection}. There are many more forms of clouds which determine the weather and provides an indication of any natural disaster, which can be handled in a low-risk manner if correctly detected beforehand.

Formation of clouds and detecting their structures and patterns beforehand allows a lot of high-risk activities to be avoided previously. Detecting the electrical discharge density of atmosphere and checking for convective patterns could reduce risk of accidents \cite{de2013new}.  Aircraft flights are a high-risk activity that carries thousands of passengers at a given interval of time, and flying through clouds is similar to driving a car through a thick fog - it is difficult to see what is ahead, making it a challenging maneuver to accomplish, as a consequence, an essential part of becoming a pilot is knowing about cloud formations \cite{zhang2011impact}. Cloud structures like \textbf{Cumulonimbus} are a direct threat to aircraft \cite{mason2006ice}. Cloud-borne updrafts and downdrafts create fast and unforeseeable outcomes to the lift force on aircraft wings which causes turbulence. These changes cause the plane to lurch and jump about during flight, known as turbulence, which sometimes makes less experienced pilots lose control of the craft, and the result is often fatal, with high casualties. Cargo ships, on the other hand, rely a lot on the weather of the sea, which is often unpredictable and ever-changing. Cargoes usually have a tight schedule and delay causes a loss of hundreds of thousands of dollars in fuel consumption. Storms also delay shipping which contributes to the loss of millions of dollars. An early prediction of storms or change in weather saves many lives and millions of dollars.

Interpretation of satellite images of cloud structures requires the expertise of a well-trained meteorologist, although it is not feasible and not always readily available. An intelligent automated system to interpret the satellite images, therefore, becomes a promising alternative with ease of access and hence becomes quite desirable for the understanding of cloud structures. 

In this paper, a model has been developed to classify the clouds into four categories using satellite images, with classification architecture like EfficientNet and segmentation architecture UNet \cite{tan2019efficientnet,ronneberger2015u}.

The research has the following contributions:
\begin{itemize}
    \item An approach capable of learning from satellite images to classify four classes of cloud shapes.
    \item Smart augmentation approach using Albumentation has been applied to expand the dataset for accurate model training.
    \item Usage of segmentation model on a classification task to show that a good encoder provides effective result with UNet as decoder.
    \item Experimental details and comparison has been provided with straightforward use of EfficientNet models (B0-B5).
\end{itemize}

\section{Dataset \label{dataset}}

The dataset consists of satellite images, courtesy of \href{https://worldview.earthdata.nasa.gov/}{ \textbf{NASA Worldview}}, gathered from Kaggle competition \href{https://www.kaggle.com/c/understanding_cloud_organization/data}{\textit{"Understanding Clouds from satellite images"}}. The images contain clouds of four classes namely: \textbf{Fish}, \textbf{Flower}, \textbf{Sugar} and \textbf{Gravel}. The images were taken from three regions, spanning 21$^\circ$ longitude and 14$^\circ$ latitude. The true color images were taken from two polar-orbiting satellites, \href{https://en.wikipedia.org/wiki/Terra_(satellite)}{Terra} and \href{https://en.wikipedia.org/wiki/Aqua_(satellite)}{Aqua}. An image might be attached from two orbits, due to the small footprint of Moderate Resolution Imaging Spectroradiometer \href{https://modis.gsfc.nasa.gov/}{\textbf{(MODIS)}} onboard these satellites. The remaining area, which has not been covered by two succeeding orbits, is marked black, as shown in \textbf{Fig. \ref{Masked}}.

The dataset was split into train and validation of 80:20. The training sample was perfectly balanced with 22184 images, which consists of $5546\times4$ images, where each class has 5546 images. All images are of the same size, which is $1400\times2100$ pixels. 68 scientists labelled images in the train set, and 3 individual ones labelled each image. Certain augmentation techniques were applied to the dataset. Augmentation is a process to artificially expand the size of the dataset by creating modified data which improves the performance of the model to generalise. Albumentations library has been used for augmentation \cite{buslaev2020albumentations}. This library efficiently implements an abundant variations in image transformation that are performance optimized. The images were augmented into four types randomly for each train data: horizontal flip, vertical flip, random 20$^\circ$ rotation and grid distortion. The count of train image data doubled after performing augmentation. Also some adjustment have been done in image size to feed into EfficentUNet architecture mentioned in \textbf{Subsection \ref{effunet}}.

\section{Materials and Methods \label{evaluation}}

\subsection{Public and Private Leaderboard Score}
Since the dataset was gathered from a public Kaggle competition, there were two sets of scores that competed among others. Public LB\footnote{LB: Leaderboard} scores are those which are shown while the competition is ongoing. It shows the outcome from a subset of the test dataset. Private LB scores get generated after the competition is over, that provides scores on the remaining test dataset. As a result, public LB scores are usually better than the private. For this particular competition, private score was calculated on 75\% of the test data. 

Pixel encoding technique was followed to participate in the submission of the competition since the image sizes were too large for Kaggle system  \cite{richards1993method}. As a result, instead of submitting an exhaustive list of indices for segmentation, pairs of values were submitted, which contained the start position and the run length of the image pixels. For example, a pair value of (1, 3) indicates that the pixel starts at 1 and run 3 pixels. The competition also required a space-delimited list of pairs. The predicted encodings were scaled by 0.25 per side, which scaled down the images of size $1400\times2100$ pixels in both train and test set to $350\times525$ pixels, hence allowing the scope to achieve reasonable submission evaluation times.

\subsection{Dice coefficient}\label{dsc}

The evaluation metric used in this paper is the Dice coefficient. It was applied to compare the pixel-wise agreement within a predicted segmentation and corresponding ground truth, using the following equation:

\begin{center}
\begin{equation}
   \frac{2*\lvert X\cap Y\rvert}{\rvert X\rvert+\rvert Y\rvert}
\end{equation}
\end{center}

Here X defines the set of pixels predicted, whereas Y defines the ground truth of the training set. When X and Y are empty, the dice coefficient is defined as 1. The LB score provides the mean of Dice coefficients for each (Image, Label) pair in the test data.

Dice coefficients are slightly different from the more popular evaluation metric: accuracy of a model. They are used to quantify the performance of image segmentation methods. Some ground truth regions are annotated in the images, and then an automated algorithm is allowed to do it. The algorithm is validated by calculating the dice score, which is a measure of how similar the objects are and is calculated by the overlap of the two segmentations divided by the total size of the two objects \cite{dice1945measures}.  Dice coefficient works better in segmentation because of its ease of differentiation, as a result it is more preferable over Jaccard's index, another evaluation metric similar to Dice coefficient \cite{niwattanakul2013using}.

\subsection{Optimizer}  
Rectified Adam (RAdam) was used instead of Adam as an optimizer for high accuracy and fewer epochs \cite{liu2019variance}.

\subsection{Loss Function}
Categorical Cross-entropy (CCE) has been applied as a loss function as it is a multi-label classification task.  To evaluate the difference between two probability distributions, Categorical Cross-entropy, also known as Softmax Loss, is used.

\begin{center}
\begin{equation}
   Loss = - \sum_{output \atop size}^{i=1} y_{i}\ .\ log\ \hat{y_{i}} 
\end{equation}
\end{center}

Here \(\hat{y}_i\) determines the \(i\)-th scalar value in the output of the model, \(y_i\) denotes the analogous target value, and number of scalar values in output of the model is provided by output size.

It's an important function in order to calculate and distinguish two probability distribution of discrete nature. \(y_i\) specifies probability that event \(i\) occurred and sum of all \(y_i\) is 1, indicating that precisely one event occurred. The function has a negative sign which confirms that the loss decreases as the distributions move closer to one another.

\textbf{Softmax} activation function is recommended with CCE, as it adjusts the output of the model and verifies if it contains the right properties, as positive outputs are desirable so that logarithm of every output value \(\hat{y}_i\) exists. Comparing two probability distribution is the major objective of this particular loss function. For the classification part loss score was calculated from the CCE and for the segmentation part it was calculated as $CCE \times 0.7 + DICE \times 0.3$.
DSC is a measure of overlap between corresponding pixel values of prediction and ground truth respectively. Range of DSC is between 0 and 1 as it is understandable from \textbf{Subsection \ref{dsc}} and the larger the better. So, Dice Loss (DICE) try to maximize the overlap between the above mentioned two sets (predcitions and ground truth pixel values) \cite{sudre2017generalised}.

\subsection{Precision-Recall (PR) Curve}

Precision-Recall (PR) curve demonstrates the relationship between positive predictive value (precision) and sensitivity (recall) of a machine learning model. Precision gives the percentage of the relevant outcome, while recall indicates the total consistent result. PR curve is a vital evaluation metric as it provides a more informative view of an algorithm's performance. X axis shows recall while the Y axis shows precision.

PR curve visualizes the set-off between recall and precision for various thresholds. A high area under the curve (AUC) shows both higher precision and recall, where they refer to a flat false-negative and low false-positive rate respectively in terms of high recall and precision. Since cloud structures are complex, it was essential to understand whether the implemented model was correctly detecting the shapes and evaluating true positive, true negative scores appropriately, hence PR curve was a crucial evaluation metric to understand the learning rate of the model.

\begin{figure}
    \centering
    \includegraphics[width=1.0\linewidth]{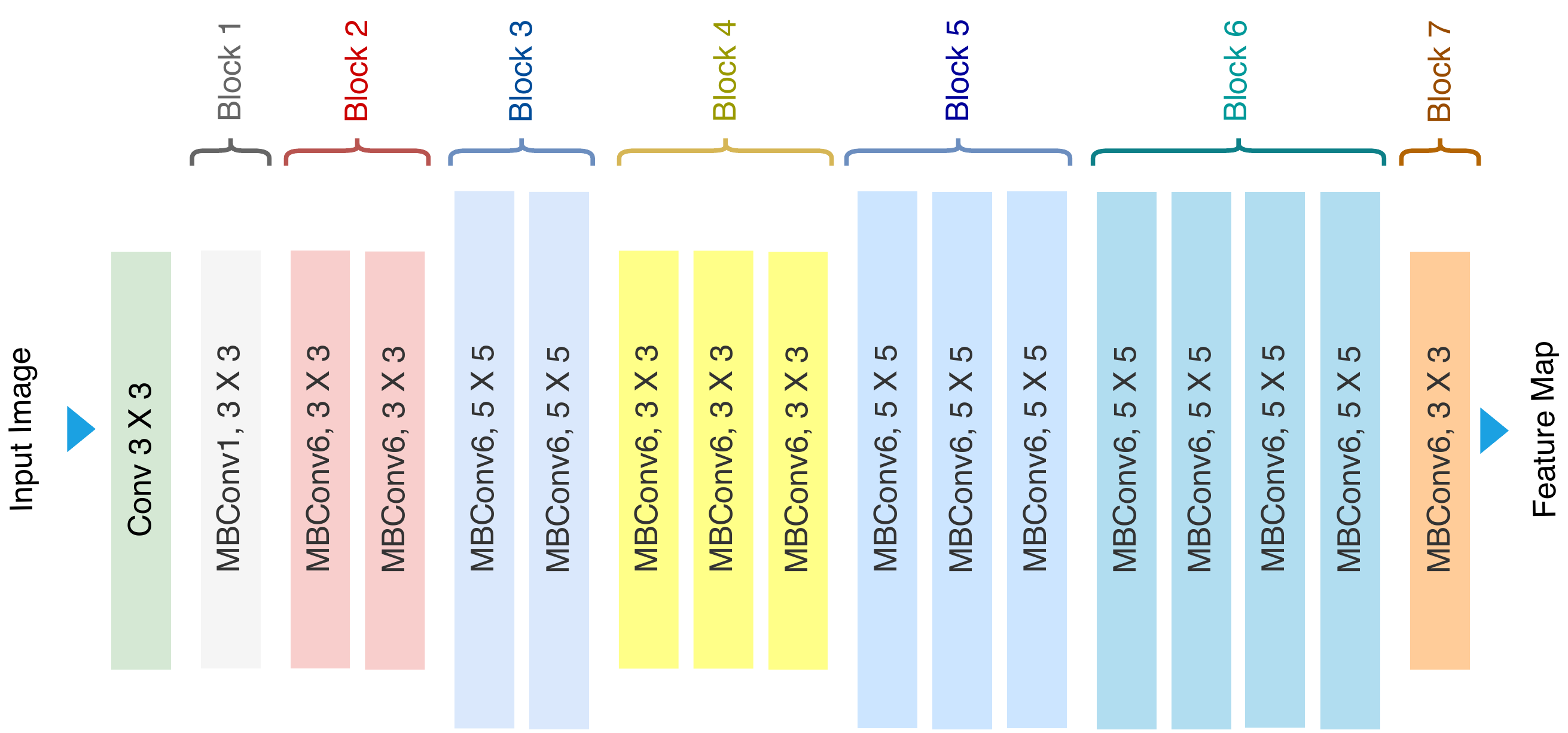}
    \caption{Architecture of EfficientNet-B0 with MBConv as Basic building blocks.}
    \label{mbconv}
\end{figure}

\subsection{Description of Applied Architectures}
\subsubsection{EfficientNet}

EfficientNet presented by \href{https://ai.google/research/}{Google AI research} is considered as a group of CNN models, but with subtle improvements, it works better than its predecessors \cite{tan2019efficientnet}. It consists of 8 variations from B0 to B7, where each subsequent model number refers to variants with more parameters and higher accuracy. EfficientNet works in three ways:
\begin{itemize}
    \item \textbf{Depthwise + Pointwise Convolution}: Depthwise convolution performs independently over each channel of input. This is a spatial convolution. Pointwise convolution projects the channel's output by the depthwise convolution onto a new channel space. This is a $1\times1$ convolution.
    \item \textbf{Inverse Res}: ResNet blocks consist of 
    \begin{itemize} 
        \item a layer that squeezes the channels
        \item a layer that extends the channels. 
    \end{itemize}
    In this way, it links skip connections to rich channel layers \cite{he2016deep}.
    \item \textbf{Linear Bottlneck}: In each block, it uses linear activation in the last layer to prevent loss of information from ReLU \cite{agarap2018deep}. 
\end{itemize}

\begin{figure}[h!]
    \centering
    \includegraphics[width=1.0\linewidth]{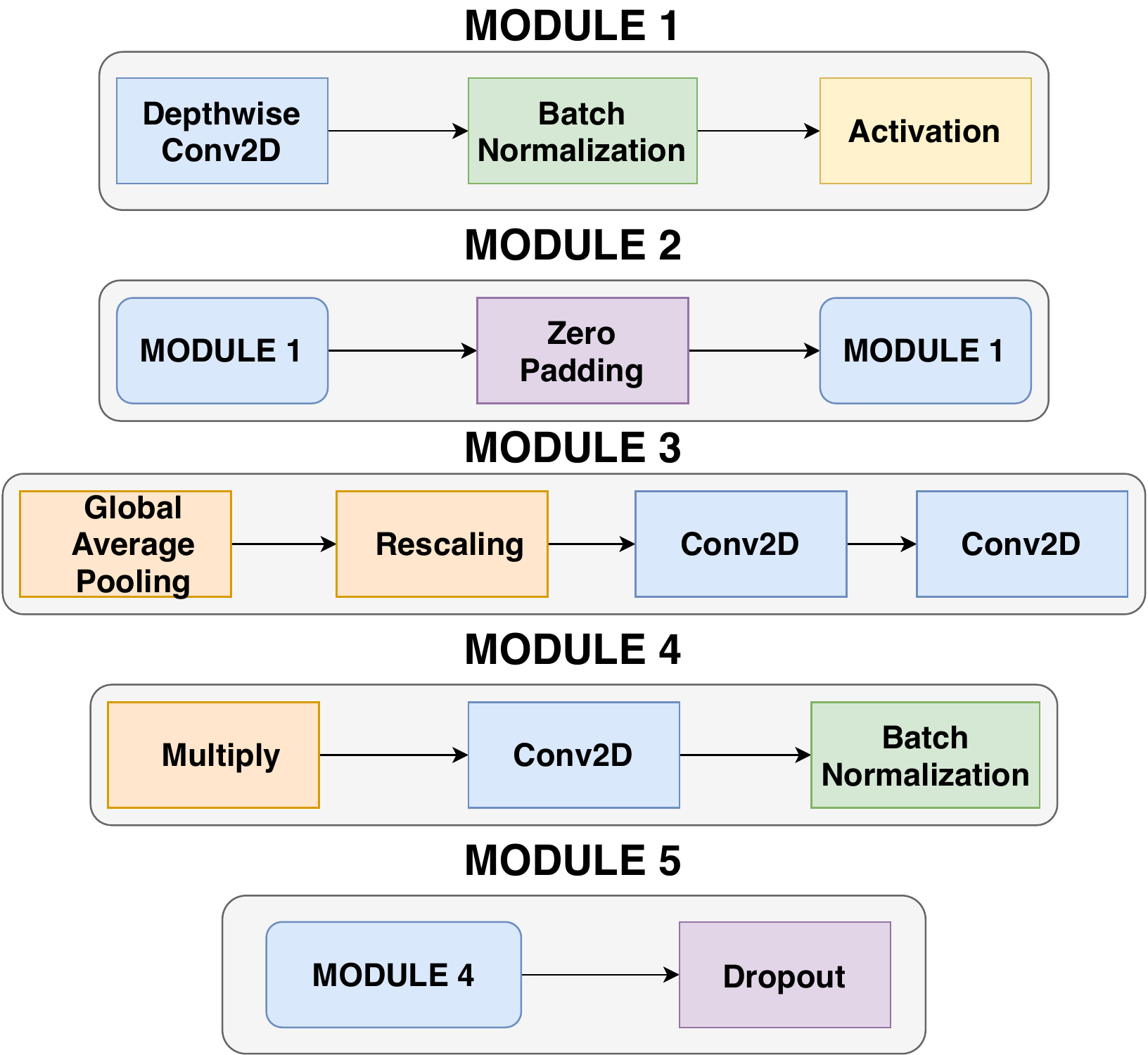}
    \caption{Common modules which were used to implement layers of all 8 models of EfficientNet}
    \label{Common_blocks}
\end{figure}

As mentioned earlier, EfficientNet has 8 variations, B0 - B7, among which, first 6 models have been explored in this paper. Due to the rise in complexity, the remaining models were ignored as they produced underappreciated results with poor performance while absorbing precious runtime. The layers in each of the models (B0 - B7) can be created by using 5 standard modules shown in \textbf{Fig. \ref{Common_blocks}}. 

\begin{itemize}
    \item \textbf{Module 1} acts as the starting block for the sub-blocks.
    \item \textbf{Module 2} acts as the initializing point for the first sub-block of all the 7 main blocks except the 1st block.
    \item \textbf{Module 3} is used as a skip connection block for all the sub-blocks.
    \item \textbf{Module 4} combines the skip connections that occurred in the first sub-blocks.
    \item \textbf{Module 5} combines each sub-block that is connected to its previous sub-block in a skip connection.
\end{itemize}

\begin{figure}[h!]
    \centering
    \includegraphics[width=1.0\linewidth]{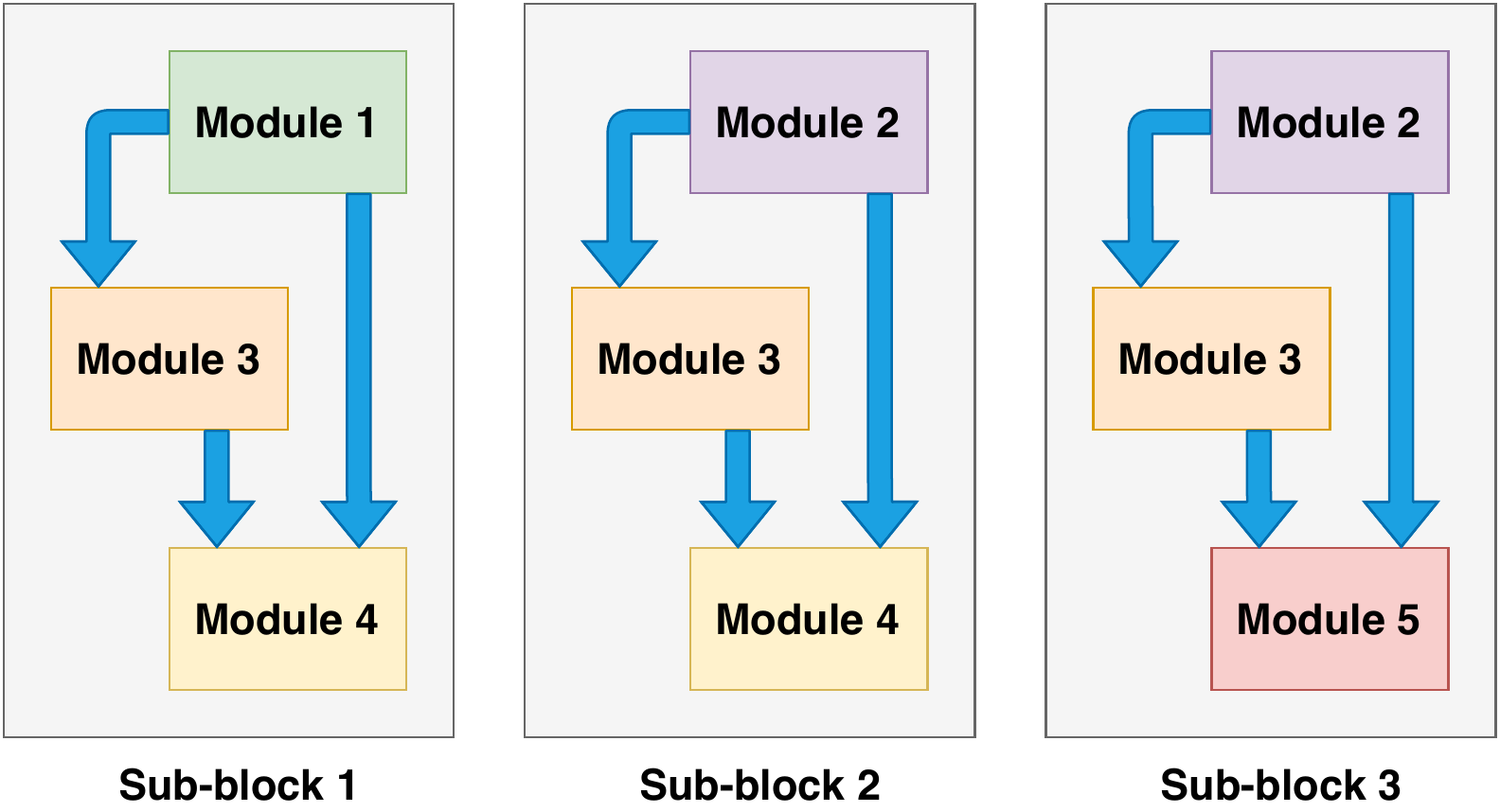}
    \caption{Sub-blocks using individual modules which are presented in figure \ref{Common_blocks}}.
    \label{submodules}
\end{figure}

The individual modules are further used in various order to create sub-blocks, as shown in \textbf{Fig. \ref{submodules}}. It’s easy to observe the difference among the models, with a gradual increase in the number of sub-blocks. The foundation for EfficientNet is MBConv layer which is an inverted residual block originally applied in MobileNetV2 \cite{sandler2018mobilenetv2}. Basic building block of EfficientNet-B0 with respect to MBConv layers have been shown in \textbf{Fig. \ref{mbconv}}. The 8 models of EfficientNet (B0 - B7) share the common blocks with subtle complexities in their architectures. 

EfficientNet is a scaled-up neural network architecture, where the models scale all dimensions with a compound coefficient, which is a newly proposed method known as \textbf{Compound Scaling} \cite{lee2020compounding}. Here scaling up is defined as a systematic, principled scaling of three factors, which are depth, width and resolution.

\begin{samepage}
\begin{itemize}
    \item \textbf{Width scale} adds more feature maps in each layer.
    \item \textbf{Depth scale} adds more layers to the network.
    \item \textbf{Resoultion scale} increase resolution of input images.
\end{itemize}
\end{samepage}

Every architecture has similarity with its earlier versions. The only difference is the different feature maps that increase the number of parameters. All the models have the same architecture as its previous one, except for the multiplied block (x2) that expands and covers more blocks. This gives a lot of parameters to be used in a calculation, making it a very robust model.
It’s not difficult to observe the changes among all the models, and they gradually increased the number of sub-blocks \cite{tan2019efficientnet}. Starting from EfficientNet-B0, compound scaling method was used to scale up with two steps:
\begin{itemize}
    \item \textbf{Step 1} coefficient was fixed to 1 assuming twice more resources to be available, and it enforces a small grid search for the networks depth, width and resolution constants.
    \item \textbf{Step 2} The constants then get fixed and scaled up the baseline network with different coefficient to obtain the successive versions from B1 to B7.
\end{itemize}

EfficientNet was prioritized in this paper due to limitations of Kaggle notebook as well. It was also the core reason why the remaining EfficientNet models were avoided, since they calculate a substantial amount of parameters, which takes a lot of processing power and time, producing a disappointing outcome. While calculating mVA for EfficientNet models, fine-tuning has been applied alongside baseline versions for each models. Imagenet pretrained model has been used while fine-tuning \cite{Yakubovskiy:2019}. 

\subsubsection{UNet}
UNet is built on the fully convolutional network. 
The structure of UNet was remodelled to work with fewer trained data which produce more accurate segmentation. The goal of UNet is enhancing a contracting layer to sequential layers. Upsampling operators are used for substituting those layers instead of pooling operations which produces better resolution of the output. A successive convolution layer learns to construct accurate output based on the data.\cite{ronneberger2015u}.

In the upsampling segment, UNet has a substantial number of feature channels which allows higher resolution layers to have propagated context information. As a result, the extensive path becomes almost symmetric to the shrinking path \cite{long2015fully}. 

The U-shaped architecture of UNet is produced by the contracting and expansive path. Contracting path is created by a typical convolutional neural network where each network is followed by a max-pooling operation and a rectified linear unit (ReLU). Throughout the contraction, spatial information is decreased while feature information is increased \cite{ronneberger2015u}.

\begin{figure*}[ht!]
 \center
  \includegraphics[width=1.0\textwidth]{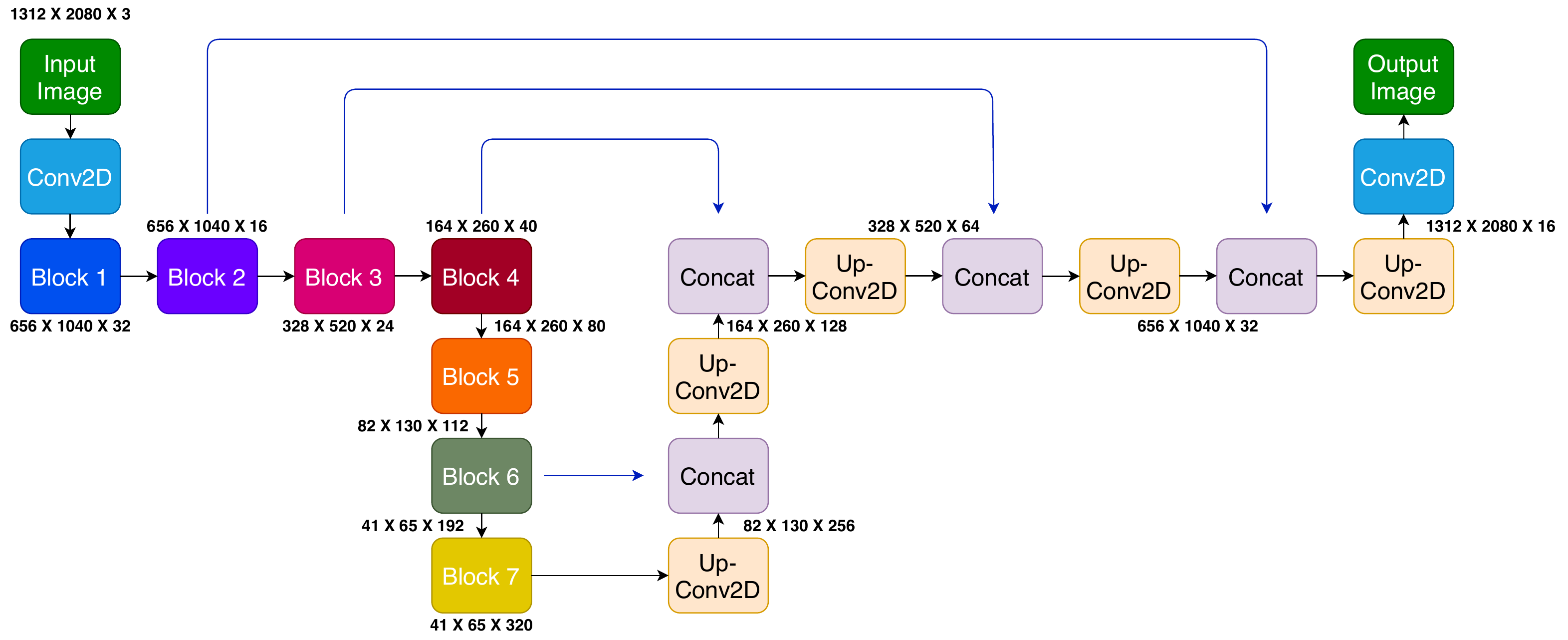}
  \caption{Architecture of EfficientUNet with EfficientNet-B0 framework for semantic segmentation. Blocks of EfficientNet-B0 as encoder has been presented in figure \ref{mbconv}}.
  \label{efficientunet}
\end{figure*}

\subsubsection{EfficientUNet}\label{effunet}
In typical UNet expansion path and the contracting path is almost symmetric. Decoder module is the same as the original UNet in this work and in the contracting path EfficientNet has been adopted as the encoder. Details of the proposed architecture including the number of channels, levels, resolution of each feature map are illustrated in \textbf{Fig. \ref{efficientunet}}. Initial size of input images is $1400\times2100$ which is resized to $1312\times2080$ for ease of processing. First, feature map of the last logit of the encoder is upsampled bi-linearly by a factor of two and then concatenated with encoders feature map maintaining the same spatial resolution. Before upsampling by the factor of two it is followed by $3\times3$ convolution layers. This process gets repeated until the segmentation map equals to the size of the input image reconstructs. Unlike the original UNet, this architecture is asymmetric where contracting path is deeper than expansion path. Addition of robust architecture like EfficientNet as encoder improved overall performance of the algorithm \cite{baheti2020eff}.

\begin{figure*}
 \center
  \includegraphics[width=1.0\linewidth]{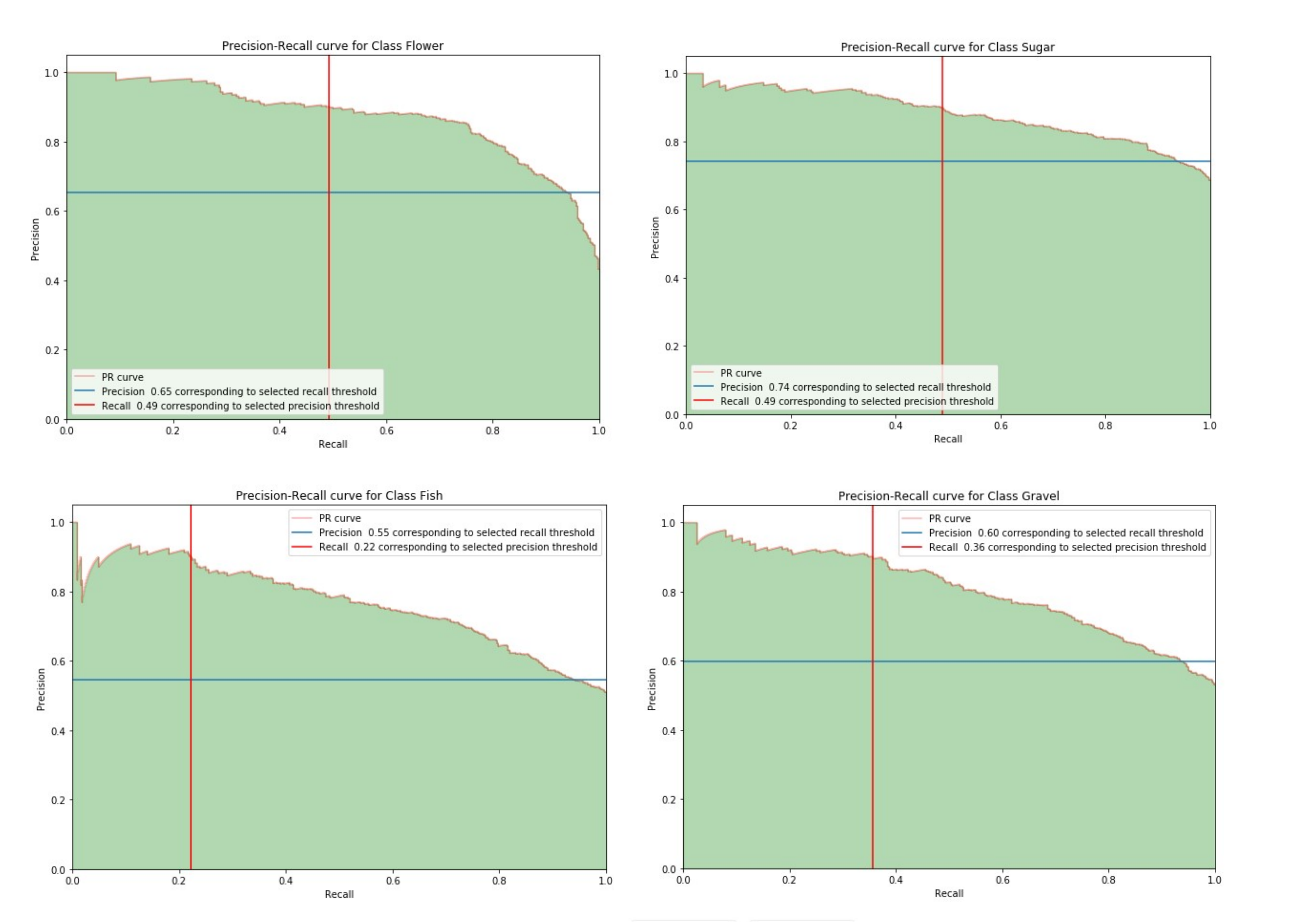}
  \caption{PR-Curves for All Four Classes}
  \label{AAA}
\end{figure*}

\section{Results and Discussion}
Experimentation has been undertaken by applying six different versions of EfficentNet architectures mentioned previously.

Mean Validation Accuracy (mVA) has been measured for all EfficentNet architectures for both baseline and fine-tuned versions.
The outcome of this part of the inspection has been presented in \textbf{Table \ref{mva}}.

\begin{table}[!ht]
\centering
\begin{tabular}{lll}
\hline
\multirow{2}{*}{\begin{tabular}[c]{@{}l@{}}EfficientNet\\ Version\end{tabular}} & \multicolumn{2}{l}{mean validation accuracy} \\ \cline{2-3} 
   & Baseline & Fine Tuned \\ \hline
B0 & 0.670    & 0.836      \\ 
B1 & 0.656    & 0.829      \\ 
B2 & 0.657    & 0.827      \\ 
B3 & 0.659    & 0.825      \\ 
B4 & 0.640    & 0.798      \\ 
B5 & 0.601    & 0.732      \\ \hline
\end{tabular}
\caption{Mean validation accuracy scores on EfficientNet version B0-B5.}
\label{mva}
\end{table}

Cross validation result along with the public and private LB score (DSC) has been presented in \textbf{Table \ref{lb}}. 

\begin{table}[!ht]
\centering
\begin{tabular}{llll}
\hline
\multirow{2}{*}{\begin{tabular}[c]{@{}l@{}}EfficientNet\\ Version\end{tabular}} &
  \multirow{2}{*}{\begin{tabular}[c]{@{}l@{}}Cross\\ Validation\end{tabular}} &
  \multicolumn{2}{l}{LB score} \\ \cline{3-4} 
   &         & Private & Public  \\ \hline
B0 & 0.6389  & 0.64595 & 0.65936 \\ 
B1 & 0.6423  & 0.64640 & 0.65849 \\ 
B2 & 0.6351 & 0.64551 & 0.65801 \\ 
B3 & 0.6311  & 0.64585 & 0.65820 \\ 
B4 & 0.6294  & 0.63911 & 0.65563 \\ 
B5 & 0.6255  & 0.64059 & 0.65325 \\ \hline
\end{tabular}
\caption{Cross Validation, Private and Public DSC leaderboard scores of EfficientNet architecture for classification.}
\label{lb}
\end{table}

The mean validation accuracy (mVA) score was best for the B0 model of EfficientNet architecture, as shown in \textbf{Table \ref{mva}}, although the other versions (B1-B5) came close on both private and public leaderboard scores, as shown in \textbf{Table \ref{lb}}. Cross validation scores were not stable since the image segmentation wasn't reliable, with fluctuations in the outcome, having only 63.89\% in B0 while the other models had a higher value. As a result it was important to use an efficient segmentation approach to reach a stable and accurate result, as shown in \textbf{Table \ref{ulb}}. 

It became visible that an improved segmentation of images stabilized the scores, with a gradual trend in the scores and model B0 showing the best response in both cross validation and LB values, with 66.54\% in cross validation, 66.02\% in private and 66.26\% in public leaderboard. The other models were not far behind either, and the scores decreased in a descending order with B1 being the second-best model with 66.01\% on cross validation, 65.53\% on private and 65.98\% on public leaderboard.

\begin{table}[!ht]
\centering
\begin{tabular}{llll}
\hline
\multirow{2}{*}{\begin{tabular}[c]{@{}l@{}}EfficientNet\\ Version\end{tabular}} &
  \multirow{2}{*}{\begin{tabular}[c]{@{}l@{}}Cross\\ Validation\end{tabular}} &
  \multicolumn{2}{c}{LB score} \\ \cline{3-4} 
   &        & Private & Public \\ \hline
B0 & 0.6689 & 0.6611  & 0.6650 \\ 
B1 & 0.6601 & 0.6553  & 0.6598 \\ 
B2 & 0.6589 & 0.6570  & 0.6578 \\ 
B3 & 0.6588 & 0.6500  & 0.6582 \\ 
B4 & 0.6425 & 0.6417  & 0.6421 \\ 
B5 & 0.6322 & 0.6319  & 0.6333 \\ \hline
\end{tabular}
\caption{Cross Validation, Private and Public DSC LB Scores of EfficientUNet for Classification}
\label{ulb}
\end{table}

Precision-Recall curves for all four classes are shown in \textbf{Fig. \ref{AAA}}. Highest precision (0.74) corresponding to recall threshold was obtained for class: Sugar and lowest for class: Fish which was 0.55, while the maximum and minimum recall for class: Flower was 0.49 and class: Fish was 0.22.

The main objective of this experiment was to express in a meaningful manner that EfficientNet alone doesn't provide a satisfying outcome in a dataset like this, despite being a substantially efficient classification layer. We showed in our research that by initiating a different segmentation architecture like UNet alongside EfficientNet, we can improve the performance on datasets like the one used in this project. It is also worth noting that instead of using UNet as a segmentation architecture alone, we used EfficientNet's encoding capability and realized that EfficientUnet shows significant improvement on the outcome. By using a classification layer as an encoder and a segmentation architecture as decoder, we boosted the performance of both the models and made them work together for a better performance.

By boosting the peformance of both the architectures, using them as an encoder and a decoder, we proved that using them in such manner allows faster performance for memory intensive datasets, and can be used in various sectors such as medical image analysis, cellular image classification for cancer detection where the data are usually segmented.

\section {Conclusion} 
This paper implements classification of satellite images of cloud structures into four different classes: Sugar, Gravel, Fish and Flower. 6 versions of EfficientNet from B0 to B5 were used as encoder and UNet as decoder has been applied. Dice coefficient was used as the evaluation metric. The scores used were compared with both public and private LB scores of Kaggle competition. By using a segmentation model like UNet in a classification problem, it was demonstrated that with a good encoder good performance can be achieved from the dataset. 

Although EfficientNet was used in this paper, it could be replaced with a different model as well but was not tested in this research. Also, a good segmentation of the images boost the output of the classification drastically, which was also demonstrated in this paper. PR-Curves were generated to demonstrate the relation between precision and recall as it shows the compromise between precision and recall for various threshold. 

In future, estimation of the distribution of classes and adjustment of the validation set could be implemented accordingly. As the complex architectures of EfficientNet gave less appreciating results because of default coefficients, altering these hyperparameters according to the dataset will hopefully improve the outcome. Exploring gradient weighted class activation mapping to generate a baseline, which is a class explainability technique, could be achieved.

\section*{Acknowledgment}
We thank \href{https://www.kaggle.com/}{Kaggle} for hosting such a great contest and for their kernel (notebook). We also thank \href{https://www.mpimet.mpg.de/en/mpimet-homepage/}{Max Planck Institute for Meteorology} for providing the dataset. We also admire \href{https://github.com/qubvel}{Pavel Yakubovskiy} for sharing his contribution in GitHub repository from where we managed to access the pretrained imagenet weights.

Source code of this project is available at \color{blue} \url{https://github.com/TashinAhmed/CloudsClassification}\color{black}.

\bibliography{bibs} 
\bibliographystyle{ieeetr}


\end{document}